\documentclass[english]{revtex4}
\usepackage[T1]{fontenc}
\usepackage[latin1]{inputenc}
\usepackage{graphicx}

\makeatletter

\usepackage{babel}
\makeatother
\begin{document}

\title{A Comprehensive Approach to Resolving the Nature of the Dark Energy}

\author{Greg Huey}

\affiliation{Department of Physics, University of Illinois, Urbana, IL 61801}

\begin{abstract}
A data-driven approach to elucidating the nature of the dark energy,
in the form of a joint analysis of a full set of cosmological parameters,
utilizing all available observational data is proposed. A parameterization
of a generalized dark energy is developed with the extension of fluid
perturbation theory to models which cross through an equation of state
of $-1$. This parameterization is selected to be general enough to
admit a wide variety of behavior, while still being physical and economical.
A Fisher matrix analysis with future high-precision CMB, cluster survey,
and SNIa data suggests the parameters will probably be resolvable
in the foreseeable future. How accurately the parameters can be determined
depends sensitively on the nature of the dark energy - particularly
how significant of a fraction of the total energy density it has been
in the past. Parameter space will be sampled at a large number of
points, with cosmological information such as CMB, power spectra,
etc of each point being archived. Thus the likelihood functions of
an arbitrary set of experiments can be applied to parameter space
with insignificant new computational cost, making a wide variety of
analyses possible. The resulting tool for Analysis and Resolution
of Dark-sector Attributes, ARDA, will be highly versatile and adaptable.
ARDA will allow the scientific community to extract parameters with
an arbitrary set of experiments and theoretical priors, test for tension
between classes of observations and investigate the effectiveness
of hypothetical experiments, while evolving in a data-driven manner.
A proof-of-concept prototype web-tool, \underbar{The Cosmic Concordance
Project}, is already available.
\end{abstract}
\maketitle

\section{Introduction}

The nature of the dark energy is one of greatest enigmas of modern
cosmology. At stake is the fate of the Universe as well as a key insight
into fundamental physics. A wide variety of theories explaining the
dark energy have been proposed - but all of them phenomenological
as opposed to fundamental. In the absence an explanation driven by
fundamental theory, our understanding must be driven by the observational
data. The coming years will provide a tremendous amount of this data,
but because the dark energy interacts weakly with the visible sector
and clumps minimally at best, elucidating its nature will be hard
work. No single set of observations can do the job alone. Successfully
determining if the dark energy is dynamical, and if so, what those
dynamics are will require the inclusion of every relevant observation
into the analysis to break parameter degeneracies, control systematics
and establish a concordance. A unified data analysis framework is
crucial. Observables that depend on the dark energy also depend on
visible-sector cosmological parameters - thus the need for a global
analysis, utilizing all data to jointly estimate all resolvable parameters.
The Analysis and Resolution of Dark-sector Attributes (ARDA) project
is such a unified framework - which will be used to estimate cosmological
parameters jointly from all relevant present and future experimental
datasets. The cosmological parameter space has been selected to allow
testing for a wide variety of classes of dark energy behavior.

The total dimensionality of the cosmological parameter space will
be large by current standards ($N\sim20$), but the parameterization
will be physically motivated, general but economical, and involve
minimal assumptions (such as stress-energy conservation) so that a
wide range of dark energy models are admitted. Those dark energy models
whose background and perturbation behavior are not contained in the
parameter space still should be detectable by looking for model signatures:
apparent disagreements between classes of observations that drive
the addition of the missing parameters (for example see~\cite{BayesianEvidence}).
Using an archive of pre-computed information, one will be able to
determine the confidence regions from an arbitrary set of experiments
- without significant new computational cost. Furthermore, one can
continue adding resolution to the likelihood function by sampling
more points indefinitely, and the parameterization can be adapted
as the incoming observational data dictates - allowing for addition
and removal of parameters without discarding work already done. Thus
the parameter space map resulting from this project will grow and
evolve with the data, never becoming redundant.

A Fisher matrix analysis suggests that these parameters may perhaps
be resolvable by foreseeable future experiments. Several fiducial
models were selected, and a joint covariance matrix was calculated
for this parameter space using CMB, supernovae luminosity distance
and cluster survey data that could conceivably be available in about
10 years. Of course it is well known that a Fisher matrix analysis
can drastically underestimate the error (confidence region), particularly
in the presence of parameter degeneracy and non-Gaussianity. However,
the result that the parameters in this case appear to be able to be
resolved well is probably reliable on a qualitative level. The parameters
are physically distinct, as are the types of experimental data. The
low level of correlation in the error may suggest there is not much
degeneracy between the parameters when data of this quality is used.
The most important factor in how much we will ultimately be able to
learn about the nature of the dark energy is how dynamical it is.
As will be shown by the Fisher matrix analysis in section 3, other
than late-time parameters $\Omega_{Q0}$ and $w_{0}$, the parameter
error will be much greater for something like a cosmological constant
than a tracking quintessence model in which the dark energy has always
been a significant fraction of the total energy density. 

With such a large dimensionality of cosmological parameter space it
is important to maximize the efficiency of the sampling algorithm
that will explore this space and determine its likelihood function.
A traditional Markov chain approach could be used, but recent results~\cite{OrigKDEIS_book,KDEIS_for_CPE}
suggest that an importance sampling with kernel density estimation
algorithm may be superior. Two advantages of importance sampling over
Markov chains is that independence of points in the former allow for
arbitrarily massive parallelism, and no step decorrelation length.

Section~2 describes the motivation for pursuing a general and comprehensive
approach to elucidating the nature of the dark energy and its physical
basis. Section~3 describes the parameter space and physical models
to be encompassed by the ARDA project. Section~4 discusses the computational
algorithms to be employed and the advantages they provide. Section~5
describes how experimental data will be modularized. Section~6 summarizes
the implications of ARDA for study of the dark energy and cosmological
parameter extraction.

\section{The Dark Energy and Underlying Physics}

A cosmological constant fits the current data well, but so do many
different models of dynamical dark energy. Even if it were determined
that the dark energy today is behaving as a cosmological constant,
it is reasonable to expect this would merely be an effective model
- in the same way that Fermi theory, with dimension-full constant
$G_{F}$, was only an effective theory of the weak interactions. We
must dig deeper to gain an understanding of the underlying physics.
Because any observation that depends on dark energy parameters necessarily
also depends on imperfectly determined parameters of the visible sector,
all cosmological parameters must be estimated simultaneously. We will
need to combine every available piece of cosmological data in a universal
joint analysis. Crucially important data will be provided by present
(WMAP~\cite{WMAP_CPE}, SDSS~\cite{SDSS_PS}, SCP~\cite{SCP_PE_latest},
etc) and future experiments (Planck~\cite{PropExp_Plank}, LSST~\cite{PropExp_LSST},
SPT~\cite{PropExp_SPT}, DES~\cite{PropExp_DES}, DUO~\cite{PropExp_DUO},
SNAP~\cite{PropExp_SNAP}, JDEM, Beyond Einstein Observatories/Probes,
etc). ARDA is an ambitious project to create a framework for joint
analysis of all cosmological data which will simultaneously yield
values for the standard cosmological parameters as well as reveal
the nature of the dark energy. The approach used will automatically
make best use of information contained in the correlation of different
types of observables (for example, CMB late ISW effect and large scale
structure evolution).

Through joint analysis of all cosmological data, the ARDA project
will lay the framework to extract the standard cosmological parameters
and answer key questions concerning the dark energy:

\begin{enumerate}
\item What is the value of the dark energy equation of state now?
\item What is the evolutionary history of the dark energy equation of state?
Did the dark energy ever track the other energy components? If so,
did it track like a scalar field?
\item Is the background evolution fully characterized by the equation of
state? or does some mechanism transfer energy to or from the dark
energy, altering its decay rate?
\item Are there measurable dark energy perturbations? If so, how have they
evolved?
\item How do these dark energy perturbations vary with scale? How are they
correlated with matter perturbations? with metric perturbations?
\item What is the sound speed and anisotropic stress of the dark energy?
\end{enumerate}
This paper describes a parameter space selected to be economical enough
that parameter values may be extracted in the foreseeable future,
general enough that a wide variety of classes of dark energy models
and types of behavior are encompassed, and physical enough that when
we do extract the dark energy parameter values they will teach us
about the (possibly rich) structure of the dark sector. Though the
experimental data that will conclusively answer the above questions
may be 10 or 20 years away, the ARDA project can begin to explore
parameter space immediately, and continue to adapt and improve indefinitely
as the volume of data grows. The ARDA exploration algorithms and parameterization
will be adapted as the data demands. Answers to the above questions
will not come easily or quickly, but a systematic joint global approach
such as what is outlined here makes maximal use of the data while
making minimal theoretical assumptions.

\section{The Theory Component: Parameter and Theory Space}

The ARDA project will allow data from current and future cosmological
experiments to be used in a joint fashion to complement one another,
constraining a concordance model of the dark energy in cosmological
parameter space. The wide range of different dark energy theories
will require a parameter space with a large number of dimensions.
The parameterization should be general, physical, allow one to smoothly
vary between different models of the dark energy, and be extensible
(that is, in the course of the analysis the removal or addition of
parameters can be done as warranted by the data). ARDA is adaptable
and open-ending - it can be improved indefinitely. As future observational
data becomes available it will incorporated, continually refining
the concordance parameter region. As the concordance model begins
to be resolved there will be a natural suggestion of additional parameters
to be added to the analysis for further refinement - this is expected
and will not render the previous work obsolete.

This project will produce a map from the space of cosmological parameters
to the space of {}``cosmological results''. The term {}``results''
means quantities that are observable or determine observables - for
example the background expansion rate as a function of redshift, linear
perturbation amplitudes of all components as a function of scale and
redshift, the cross correlation between those perturbations, the cosmic
microwave background anisotropy (CMB), and the likelihood with respect
to select experimental data. Such results can be used to predict experiment
observations (exp: supernovae luminosity distances, matter power spectrum,
etc) as a function of point in parameter space. Thus with little new
computational cost one can transform an arbitrary set of observations
into a likelihood function over parameter space. One can find bounds
imposed by all of the data, or by subclasses of the data (high or
low z, large or small scale, etc). There are well-defined statistical
tests that can detect if subclasses of the data produce inconsistent
parameter bounds. For example, if one assumed the dark energy equation
of state $w$ was constant, but then found high-z data favored a value
inconsistent with that favored by low-z data, one would be forced
to change that assumption and introduce a parameter to describe the
evolution of $w$.

\subsection{The Parameter Space}

The ARDA parameter space has been chosen according to the following
considerations:

\begin{enumerate}
\item The parameters describing the dark energy must be general enough to
encompass the widest range of models that is practical. The parameter
space may include a model as a subspace, or instead it may permit
tests for signatures of a model. Minimal assumptions are made with
respect to the dark energy - such as stress-energy conservation.
\item One should be able to continuously vary the cosmology between dark
energy models. An allowed region will ultimately bound the correct
model as a concordance between multiple experimental data sets.
\item The parameters should be physically motivated. Physical motivation
is important to avoid unphysical parameter combinations, and additionally
when a parameter is measured one will learn something about the underlying
physics.
\item The parameters space should be economical. A balance must be kept
between having a parameter space general enough to contain or allow
tests for signatures of the maximum variety of dark energy models,
while keeping it small enough enough so that one can obtain meaningful
parameter bounds from observable data in the foreseeable future. The
solution is a data-driven approach: allow enough parameters to resolve
general classes of effects as the data becomes good enough, but not
to allow for time or scale variation of quantities when we can not
yet determine their average values. However, as improved data suggests
additional parameters, they are added - a new dimension opens in parameter
space. 
\end{enumerate}
Note that in the assumption of conservation of the dark energy stress-energy
tensor, interactions with other components of the Universe is not
allowed for. However, one can compare a dark energy (matter) decay
rate determined from data on background evolution with dark energy
(matter) pressure determined from data on perturbation growth. An
inconsistency would be a signature for stress-energy non-conservation
of a component - energy transfer through an interaction. Although
dark energy perturbation data will be weak at best, evidence for dark
energy-matter interactions may be extracted from dark matter perturbations.
One may also treat interacting dark energy-matter as a single fluid,
which would then have significant perturbations.

The cosmological parameters to be used in ARDA are as follows:

General cosmological parameters: Hubble constant ($h$), baryon density
($\Omega_{B}h^{2}$), cold dark matter density ($\Omega_{C}h^{2}$),
neutrino energy density ($\Omega_{\upsilon}h^{2}$), reionization
optical depth ($\tau$), Helium-4 density ($Y_{p}$), number of massless
neutrinos ($N_{\upsilon}$), number of massive neutrinos ($N_{M\upsilon}$),
spatial curvature ($\Omega_{K}$), primordial scalar perturbation
amplitude ($A$), scalar perturbation spectral index ($n_{s}$), tensor
spectral index ($n_{t}$), tensor to scalar perturbation ratio ($r$),
scalar index running with scale ($d\ln n_{s}/d\ln k$), tensor index
running with scale ($d\ln n_{t}/d\ln k$), 

Dark Energy parameters:

\begin{enumerate}
\item Fraction of critical density in dark energy today: $\Omega_{D}\equiv\frac{\rho_{D}}{\rho_{tot}}\equiv1-\Omega_{K}-\Omega_{C}-\Omega_{B}-\Omega_{\nu}$
(dependent parameter)
\item Equation of state today: $w_{0}$ ($w_{0}$ may be >, = or < -1)
\item Number of expansion e-folds since tracking behavior ceased, and $w$
started to change to present value: $N_{a}=\ln\left(z_{a}+1\right)$.
Note that $w$ may pass through $-1$.
\item Rate of transition of $w$ from the tracking value to $w_{0}$: $s\equiv\frac{dw}{dN}=const$
\item Log of the ratio of the dark energy fraction during radiation domination
to the fraction during matter domination: $\mu_{R}$
\item Dark energy rest-frame sound speed: $c_{s}^{2}$
\item Dark energy anisotropic stress: $\Sigma$
\end{enumerate}
The parameter space will be modified in a data-driven fashion. For
example, we allow the dark energy perturbation rest-frame sound speed
($c_{s}^{2}$) to vary away from unity (allowing non-scalar field
dark energy) but do not allow it to vary with time or scale. If in
the future the data is good enough to resolve $c_{s}^{2}$, one may
observe a disagreement in the favored value between early and late-time
data (there are well-established statistical tests for such a purpose).
One then has reason to add new parameters that describe the variation
of $c_{s}^{2}$ with time, and the data will good enough to start
to resolve those parameters. Furthermore, tracking behavior for the
dark energy is allowed, but not required. It may be eliminated by
making $N_{a}$ large enough so that dark energy fraction before $N_{a}$
is negligible. Tracking does not assume the dark energy is a scalar
field, as $c_{s}^{2}$ and $\mu_{R}$ are allowed to vary away from
scalar field values ($1$, $\ln\left(4/3\right)$), and $w_{0}$ may
be $<-1$. 

The following are examples of degrees of freedom not present in the
parameter space, but may be added when and if called for by the observational
data:

\begin{enumerate}
\item Variation of $c_{s}^{2}$, $\Sigma$ with scale or time.
\item Non-adiabatic initial conditions
\item Arbitrary splined time dependence of $w$
\item Measurable non-tracking behavior of the dark energy density before
$N_{a}$. A standard cosmological constant may be specified, for example,
by taking $N_{a}$ large.
\end{enumerate}

\subsection{Testing for Dark Energy Models}

The following list is a sample of dark energy models that can be tested
for within the framework of this parameter space. If their background
and perturbation behavior are contained within the parameter space
then the test is direct. Otherwise one may test for signatures of
those types of behavior not explicitly contained.

\begin{enumerate}
\item cosmological constant (constant $w=-1$)
\item simplified quintessence (constant $w>-1$)
\item simplified quintessence (like above, but with a constant parameter
describing the time variation of $w$, $w_{0}>-1$)
\item general tracking quintessence~\cite{DE_trakQEarly,DE_trakQPJS}:
dark energy to background energy density ratio $\rho_{DE}/\rho_{BG}\sim constant$
until $N_{a}$, then $w$ changes to present value $w_{0}$. {}``general''
means that a scalar field is not assumed.
\item general creeping quintessence~\cite{DE_crepQ_CantFindPJSRef}: A
dynamical degree of freedom is mimicking a cosmological constant in
terms of background behavior now and in the recent past. Because it
is dynamical it may have perturbations.
\item Hybrid quintessence potentials: With varying degrees of naturalness,
a more complicated scalar field potential yields tracking at high
energy and quintessence domination at late times when the field encounters
a feature in the potential (for example Albrecht \& Skordis constructed
a model using only Planck-scale parameters where a potential is exponential
at high energy, but has a low energy minimum~\cite{DE_AlbrechtSkordis}).
These models track at early times and have $w_{0}\simeq-1$, $\mu_{R}=4/3$,
$c_{s}^{2}=1$, $\Sigma=0$.
\item Phantom dark energy ($w_{0}<-1$)~\cite{DE_Phantom}: The underlying
physics of the dark energy causes it to possess or develop a current
equation of state $w_{0}<-1$. This violates the dominant energy condition,
but perturbation evolution in such models is not necessarily pathological.
\item Cardassian model modifications to General Relativity~\cite{DE_Card_Orig}:
The background is modeled by a possibly time-varying $w$, large $N_{a}$
(it is a non-tracker). Perturbation evolution is modeled by a non-unity
$c_{s}^{2}$ and zero $\Sigma$~\cite{DE_Card_Fluid}.
\item Vacuum-driven Metamorphosis~\cite{DE_Metamorph}: Non-perturbative
physics of the vacuum cause a transition to a cosmology that will
eventually resemble a cosmological constant with continuous production
of radiation. The background would evolve as dark energy with a $w$
that has only recently (redshift$\sim1$) become $-1$.
\item Quantum effects of a massless, minimally coupled scalar field with
quartic self-interaction can cause the field to develop an average
equation of state $w_{0}<-1$~\cite{OnemliWoodard}. However, current
estimates of the magnitude of $\left|w_{0}+1\right|$ suggest the
background evolution of such a model may be indistinguishable from
a cosmological constant. It is not yet clear how perturbations would
behave - this may provide a critical means of distinguishability.
\item Interacting quintessence - dark matter models~\cite{DE_iQCDM,DE_otheriQCDM}:
The quintessence field interacts with one component of the dark matter,
causing it to decay at a slower rate. The interaction also suppresses
the kinetic energy of the quintessence field, causing it to drive
cosmic acceleration. Because a component of the dark matter clumps
like matter but decays like quintessence, one signature of such a
model is non-concordance of CMB, supernovae and large scale structure
data in terms of the parameters ($\Omega_{m}$, $w$) (that is, if
one makes the mistaken assumption that everything that clumps like
matter decays like matter, then one finds there can be no agreement
between the above three classes of observations).
\end{enumerate}

\subsection{Evolving the Dark Energy}

An general way to model the dark energy fluid is as a general, non-interacting
fluid that is covariantly conserved within the framework of general
relativity. The background density evolves according to\begin{equation}
\frac{d\ln\rho_{D}}{dN}=-3\left(w+1\right)\label{eq:BGevolv}\end{equation}
while the perturbations evolve as\begin{equation}
\begin{array}{c}
\dot{\delta}=-3\frac{\dot{a}}{a}\left(\widetilde{c_{s}^{2}}-w\right)\delta-\zeta-\left(1+w\right)\dot{\Phi}\\
\dot{\zeta}=-3\frac{\dot{a}}{a}\left(\frac{1}{3}-w\right)\zeta+k^{2}\widetilde{c_{s}^{2}}\delta-k^{2}\left(w+1\right)\sigma+k^{2}\left(w+1\right)\Psi\end{array}\label{eq:PertEvolv}\end{equation}
where $\delta=\delta\rho_{D}/\rho_{D}$ is the energy density perturbation,
$\zeta\equiv\theta_{D}\left(w+1\right)=ik\cdot v_{D}\left(w+1\right)$
is the momentum perturbation, in (synchronous, conformal) gauges the
metric perturbations $\Phi=\left(h/2,-3\phi\right),$$\Psi=\left(0,\psi\right)$,
and the dot denotes a derivative with respect to conformal time. These
perturbation equations are based on those appearing in~\cite{PertEvol2_MB},
but modified to allow $w$ to vary from $w>-1$ to $w\leq-1$ smoothly.
The symbols $\delta,\theta,\sigma,h,\phi,\psi$ have the same definitions
as in~\cite{PertEvol2_MB}. For numerical convenience the perturbations
are usually evolved in the dark matter rest frame. The dark energy
sound speed $\widetilde{c_{s}^{2}}$ in terms of its value in the
dark energy rest frame $c_{s}^{2}$ is approximately~\cite{Cs2_xform}:\begin{equation}
\widetilde{c_{s}^{2}}\delta=c_{s}^{2}\delta+\left[3\frac{\dot{a}}{a}\frac{\zeta}{k}\left(c_{s}^{2}-w\right)+\frac{dw}{dN}I\left(w\right)\right]\label{eq:CS2_xform}\end{equation}
This is nothing more than a frame transformation. It is the quantity
$c_{s}^{2}$ that we take to be a constant parameter, independent
of time and scale. Normally one would take $I\left(w\right)\equiv\left(w+1\right)^{-1}$.
However, as a changing $w$ passes through $-1$ the velocity perturbation
$\theta$ associated with a finite $\zeta$ diverges, and the sound
speed transform equation becomes invalid. Physically, there is nothing
pathological - $\delta$ and $\zeta$ remain finite. Thus we impose
an ad-hoc cutoff $\varepsilon$ in $I\left(w\right)$:\begin{equation}
I\left(w\right)\equiv\left\{ \begin{array}{c}
\left|w+1\right|>\varepsilon\::\:\frac{1}{w+1}\\
\left|w+1\right|<\varepsilon\::\:\frac{w+1}{\varepsilon^{2}}\end{array}\right\} \label{eq:cs2_xcutoff}\end{equation}
Testing with several sample models show no observable effect as $\varepsilon$
is varied around $0.01$. However, a more sophisticated prescription
for determining an appropriate value of $\varepsilon$ may be developed
as needed. The reason the sound speed transformation goes singular
as $w$ passes through $-1$ is that the dark energy density has reached
a minimum value and will start increasing with expansion. One does
not have freedom one otherwise would to decrease the energy density
perturbation at given coordinate value on a constant-time hypersurface
by deforming that hypersurface point to a later time. Thus eq~\ref{eq:CS2_xform}
is valid outside of a small interval around $w=-1$. Eq~\ref{eq:cs2_xcutoff}
is a way of replacing the behavior of the transform inside this interval
with an interpolation that matches the transform outside of the interval.

As with the sound speed $\widetilde{c_{s}^{2}}$, without explicit
Euler-Lagrange equations of motion for the dark energy there is some
freedom in specifying the anisotropic stress $\sigma$. One possible
choice is to assign a scale and time dependence to $\sigma$ that
is consistent with viscous damping of velocity perturbations in shear-free
frames. The dependence is further specified by imposing a special
boundary condition on the highest mode in the angular moment hierarchy~\cite{GenDarkMatter}.
This boundary condition - inspired by photons and neutrinos - allows
the angular moment hierarchy to be truncated at the quadrupole in
numerical simulations. We adopt a slightly modified form of this relation,
which allows the perturbation evolution to be continued into the regime
where $w<-1$:\begin{equation}
\dot{\xi}+3\frac{\dot{a}}{a}\xi-\frac{\dot{w}}{w}\xi=\frac{8}{3}\Sigma\left(\zeta-\left(w+1\right)\dot{\Upsilon}\right)\label{eq:AnisotropcStressPrescptn}\end{equation}
where $\xi\equiv\sigma\left(w+1\right)$, $\Sigma$ is the constant
parameter that was above called anisotropic stress and in the (synchronous,
conformal) gauges the metric perturbation $\Upsilon=\left(-h/2-3\eta,0\right)$.
In the limit $w=0$ the above relation is identical to equation~12
of~\cite{GenDarkMatter}. This would apply if the dark energy is
tracking during matter-domination. When $w=const\neq0$ the above
is equivalent to a rescaling of $c_{vis}^{2}$ of~\cite{GenDarkMatter},
changing the sign for $w<-1$. Different choices for the anisotropic
stress certainly are possible, but any alternate prescription should
preserve the physical effect of damping of velocity perturbations
for $\Sigma>0$.

A point worth noting is that these equations show a fluid with $w=-1$
can have density and momentum perturbations - they simply are not
sourced by metric perturbations. The perturbations may be primordial,
or may be sourced in an earlier epoch before $w\rightarrow-1$. In
the case of scalar field quintessence $w$ may go to $-1$ at late
times, for example, if the potential becomes flat or the field is
caught in a minimum. These decoupled perturbations then decay away,
but could perhaps be measurable today. In such a way we might hope
to distinguish between a fundamental and effective cosmological constant.

\subsection{Fisher Matrix Analysis with Future High-Precision Experiments}

A Fisher matrix analysis suggests that the ARDA cosmological parameters
may be resolvable by foreseeable future experiments. Five flat fiducial
models consistent with current CMB (WMAP~\cite{WMAP_CPE} + other
high-$\ell$ data), Sloan~\cite{SDSS_PS} and Supernovae~\cite{SCP_PE_latest}
data were selected:

\begin{enumerate}
\item Cosmological constant: $w_{0}=-1.00$, $\Omega_{D}=0.70$
\item Tracking quintessence: radiation domination: $w\simeq\frac{1}{3}$,
$\rho_{D}\sim0.12\rho_{tot}$, matter domination: $w\simeq0$, $\rho_{D}\sim0.09\rho_{tot}$,
transition to dark energy domination now: $w_{0}=-0.95$, $\Omega_{D}=0.73$
\item Phantom dark energy: $w_{0}=-1.05$, $\Omega_{D}=0.74$
\item Tracking quintessence: radiation domination: $w\simeq\frac{1}{3}$,
$\rho_{D}\sim0.15\rho_{tot}$, matter domination: $w\simeq0$, $\rho_{D}\sim0.12\rho_{tot}$,
transition to dark energy domination now as an effective cosmological
constant: $w_{0}=-1.00$, $\Omega_{D}=0.73$
\item Tracking Phantom dark energy: radiation domination: $w\simeq\frac{1}{3}$,
$\rho_{D}\sim0.15\rho_{tot}$, matter domination: $w\simeq0$, $\rho_{D}\sim0.12\rho_{tot}$,
transition to Phantom dark energy domination now $w=-1.05$, $\Omega_{D}=0.73$
at late times
\end{enumerate}
All of the tracking models have $c_{s}^{2}=0.75$, $\Sigma=0$ and
$\mu_{R}=\ln\left(4/3\right)$. None of these fiducial models have
any interactions (other than gravitational) within the dark sector
or between the dark and visible sectors. Models 1 and 3 tracked at
very early times, but then ceased tracking at $z>50$ and then begin
to behave as a cosmological constant or Phantom energy respectively.
While tracking, the fractional energy density of the dark energy ($\sim10^{-5}$)
is so small as to make the models effectively a cosmological constant
or Phantom energy with respect to observations.

A joint covariance matrix was calculated for this parameter space
via Fisher matrix analysis using data that could conceivably become
available within about 10 years:

\begin{enumerate}
\item CMB: $\ell_{max}(TT,EE,BB,ET)=2000,2000,1000,2000$, cosmic variance
of $\frac{2}{3}$sky with a minimum noise contribution to the $C_{\ell}$
of $10^{-3}\mu K^{2}$
\item supernovae luminosity distance: 8000, 3\% $d_{L}H_{0}$ error, $z_{max}=1.7$
\item Cluster survey: volume-limited, analyzed as $14$ independent redshift
bands out to $z_{max}=2.9$ as discussed in~\cite{zRingsPS}
\end{enumerate}
As can be seen in figure~1, other than late-time parameters $\Omega_{Q0}$
and $w_{0}$, the error is much greater for something like a cosmological
constant than a tracking quintessence model in which the dark energy
has always been a significant fraction. For the dark energy to be
a significant fraction in the early Universe a dynamical model such
as tracking quintessence seems likely. For this reason, the more similar
the dark energy is to a non-dynamical cosmological constant, the less
we will learn about the underlying physics. Our ultimate success in
understanding the physical nature of the dark energy will most directly
depend on how fortunate we are.

Of course the usual Fisher matrix caveats apply - the true confidence
region can be significantly underestimated, particularly in the presence
of parameter degeneracy and non-Gaussianity. However, it should be
noted that this analysis treated the experiments as independent and
hence discarded information from the correlation between the experiments
- ARDA will automatically incorporate this information. Still, the
proper way to determine the confidence region would be to perform
an analysis such as what ARDA will ultimately do - were such a result
available now something like ARDA would already have been completed.
However, the three types of experimental data are the product of very
different physical processes and the data is very precise, leading
one to expect they are unlikely to share degeneracies. Additionally,
the parameters are very distinct physically. Within the tracking models,
for which the parameters are best resolved, the Fisher matrix analysis
indicates a low level of correlation between parameters, which may
suggest there is not much degeneracy. Though one should not trust
the Fisher matrix results quantitatively, it is probably qualitatively
reliable that the parameters will ultimately be accurately resolved
- particularly for tracking models.

\begin{figure}
\includegraphics[%
  scale=0.3]{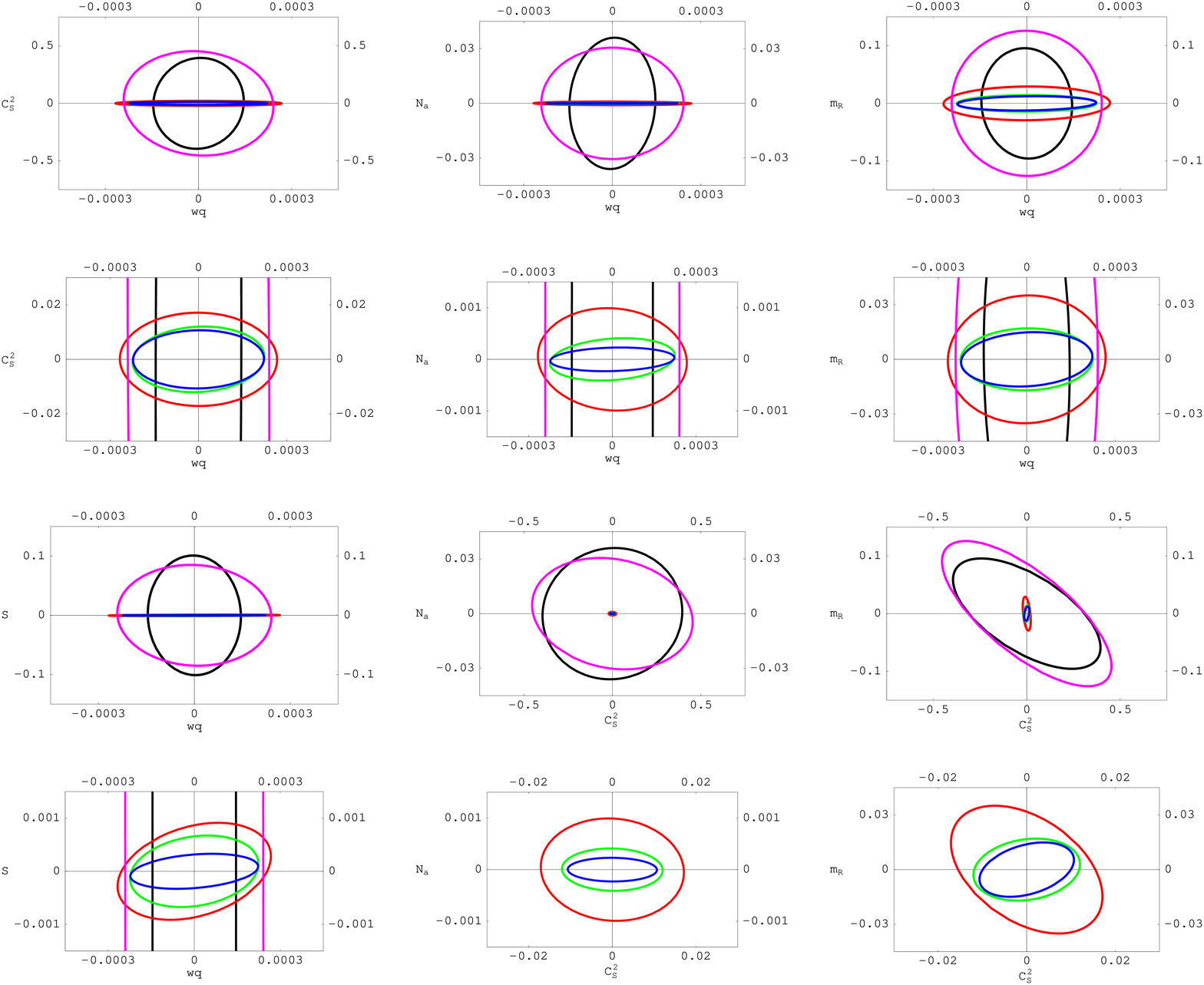}

Figure 1: Shown are the Fisher Matrix 1$\sigma$ error ellipses (with
means subtracted) of the independent dark energy parameters. The hypothetical
observational data is CMB, deep volume-limited cluster survey power
spectra, and supernovae luminosity-distances. The cosmological constant
model is black, phantom energy violet, tracking with $w_{0}=-0.95$
red, tracking with $w_{0}=-1.0$ green, and tracking with $w_{0}=-1.05$
blue. The parameter symbols are as follow: $wq$ is current dark energy
equation of state, $N_{a}$ is the number of expansion e-folds since
tracking ceased, $s$ is the rate of change of the dark energy equation
of state to its current value ($\frac{dw}{dN}$), $c_{S}^{2}$is the
dark energy sound speed, and $\mu_{R}$is the log of the radiation
domination dark energy density fraction multiple.
\end{figure}
\begin{figure}
\includegraphics[%
  scale=0.3]{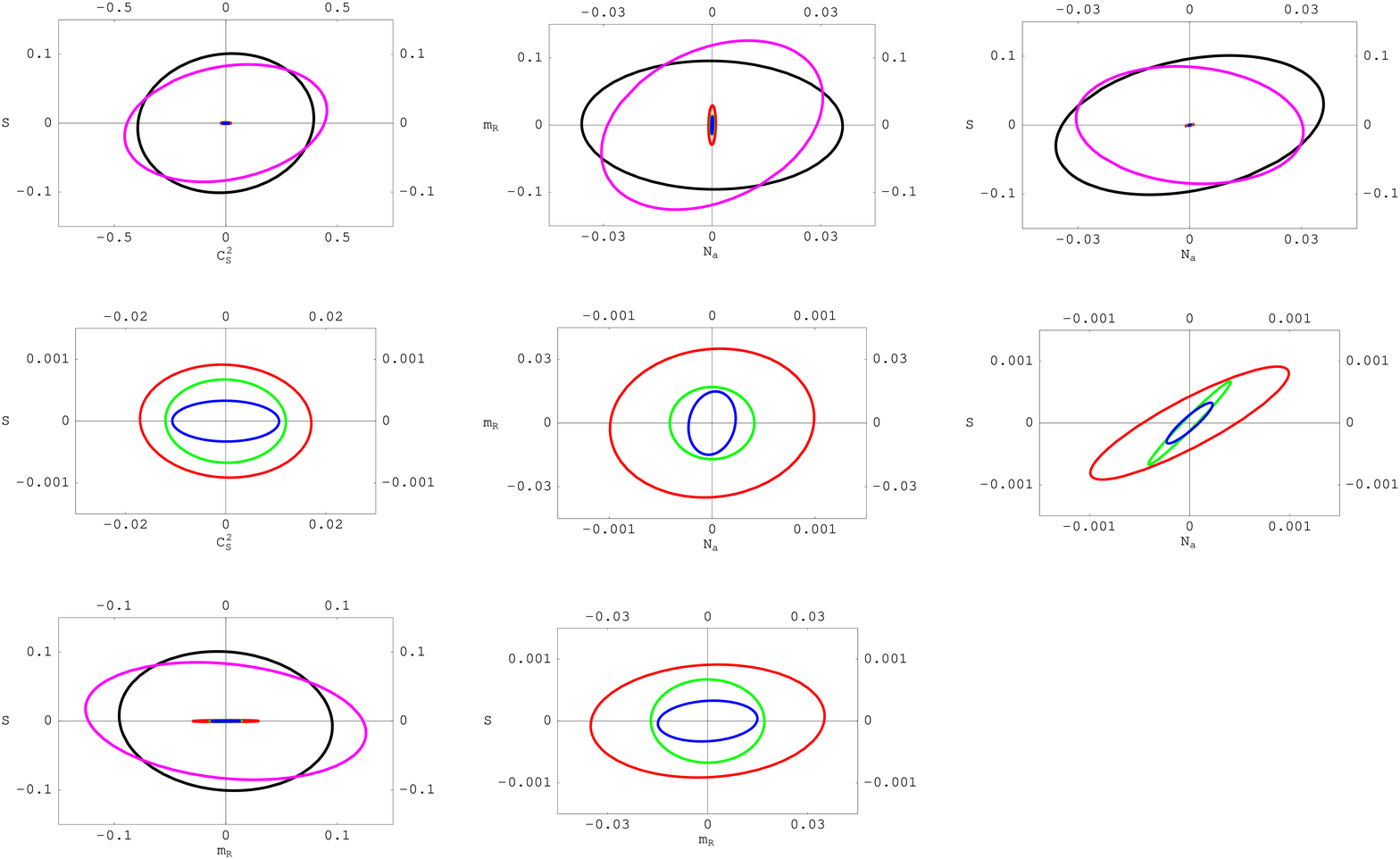}

Figure 1 (cont)
\end{figure}

\section{The Computational Component: Algorithms and CPU Sources}

The likelihood function in the $21$ dimensional parameter space will
be constructed by sampling the space over a large number of points.
A Markov chain sampling algorithm could be used for this purpose.
Recent results - described in detail in~\cite{KDEIS_for_CPE} - suggest
that importance sampling may be a superior algorithm for cosmological
parameter estimation. Importance Sampling draws a large number of
points from an (estimated) proposal distribution, and calculates the
actual (unnormalized) likelihood at each point. The proposal distribution
for the next iteration is constructed by Kernel Density Estimation
(KDE) as a sum of Gaussian distributions (kernels) - each point of
the previous iteration is replaced by a Gaussian kernel of weight
determined by the ratio of the previous proposed and actual likelihoods.
A convergence test is the correlation between the proposed and actual
likelihoods. Thus ultimately the underlying distribution in parameter
space is approximated by a large number of overlapping Gaussian distributions.
The final proposal likelihood function agrees with the underlying
likelihood, and the fact that it is a smooth and continuous function
is very useful for further analysis. Even if a Markov chain algorithm
is used for sampling, Kernel Density Estimation will still be used
for analysis purposes.

In the case of ARDA, the underlying distribution - hereafter referred
to as the ARDA base distribution - will be that of recent CMB experiments
(WMAP and others). The current plan is that no non-CMB data will used
for the base distribution. For each point the following will be archived:
the parameter values, proposed \& actual likelihoods, the CMB spectra,
and density power spectra \& transfer functions of all components
over a wide range of redshift and scale. Thus for each parameter space
point, a hypothetical observation for a given experiment can be computed
later with minimal cost. The set of parameter space points will converge
to the likelihood function of the current CMB data. This likelihood
can be multiplied by the likelihood function of any arbitrary set
of other cosmological experiments. Conceptually, this can be done
by re-weighting the parameter space points or kernels - each point
is assigned a weight equal to the product of its likelihood in each
of the set of experiments. In this way any set of experiments can
be added as if they were a prior, but after the fact. It is important
to be able to find the confidence regions in parameter space for many
different combinations of observational datasets. One reason is to
test for disagreement between large vs. small scale observational
data, and high vs. low z - an indication that a presumed constant
parameter in fact varies. Thus it is desirable to compute the ARDA
base likelihood distribution with the minimal amount of data that
will still result in an accurate resolution of the structure of the
likelihood function. Removal of a dataset from the base distribution
(by necessity through deconvolution - that is, positive re-weighting)
should be avoided because the tails of the distribution may not be
as well covered (sampled). Thus, experiments used in the base distribution
should be those that one would rarely - preferably never - want to
remove. However, the base distribution should be sufficiently constrained
so that it is well sampled - particularly the degeneracies need to
be resolved well. For ARDA the choice of CMB-only data as the data
determining the base distribution is thought to be fairly optimal,
but as the project develops this presumption will be tested and changed
if necessary.

As an example of the potential of alternate methods for exploring
parameter space, consider Importance Sampling algorithms. With Importance
Sampling the points drawn from the proposal distribution in a single
iteration are statistically independent - meaning they may be computed
massively in parallel. In recent testing, with surplus time on a retiring
NCSA cluster (200-600 Pentium IIIs used at a time), it was shown that
rates in excess of $10^{5}$ points per day are possible. As a comparison,
the WMAP CMB-only parameter extraction was done with somewhat more
than 30,000 Markov chain points~\cite{WMAP_PEMCMC}. A Markov chain
must proceed sequentially because the coordinates of the next point
can not be determined until the likelihood of the current point has
been computed. A contingency set of points can be simultaneously precomputed,
but this is inefficient: for $n$ steps of precomputation $2\left(2^{n}-1\right)$
points are computed, $n$ are used and the rest discarded. With an
Importance Sampling algorithm, massive numbers of cheap, slow computers
can be used for unlimited parallelism. Additionally, there is no correlation
length in the point sample. Depending on the dimensionality and other
factors, cosmological parameter estimation Markov chains typically
need to be thinned by a factor of order $100$ to obtain a truly uncorrelated
sample of the underlying distribution - yielding $1$\% efficiency.
Importance Sampling does not have this problem because the entire
sample is drawn from a distribution that does not depend on any of
the points in that sample. Further investigation may show that these
features, coupled with the availability of CPU time on cheap and slow
computers, expedite the convergence of the ARDA base distribution
and thus are significant advantages of Importance Sampling over Markov
chains. However, if no sampling algorithm can be found that yields
convergence of the ARDA base distribution in a reasonable amount of
time, additional observational data can be used along with the CMB
to reduce parameter degeneracy and shrink confidence regions. A disadvantage
is that the additional prior may not be able to be removed (ie: deconvolved)
while keeping good point coverage of parameter space with a fixed
number of points (were this possible, one would not need the additional
prior). Initially this project will primarily use standard dedicated
supercomputer time. This may be supplemented by CPU from otherwise
idle workstations. A server-client model is being developed that will
allow a user to donate their idle workstation CPU time to this project
by running the client in the background - much in the same way as
SETI@Home works. Though it is not relied upon, there is a potential
to tap a vast quantity of unused CPU time. Such a scenario might not
be feasible if the sampling algorithm is a Markov chain, however it
would be ideally suited to Importance Sampling.

A principle concern with such a large dimensionality of parameter
space is whether the sampling algorithm can resolve degeneracies in
the base distribution. There will be parameter space degeneracies
in the CMB - the concern is that the degeneracies might not be {}``found''
or sufficiently well sampled by the sampling algorithm due to them
having a large volume and complicated structure. Note that a poorly
constrained but uncorrelated parameter is not a problem - this will
simply result in the kernel size being large in this direction, and
resolution of the other parameters will not be degraded. Clearly a
definite answer to this degeneracy issue requires knowing the nature
and extent of the parameter degeneracies - which in turn requires
that one has fully explored the likelihood function of current CMB
experiments in this parameter space. Thus, at the beginning of the
ARDA project there is no definite answer. However, one can pick physically
distinct parameters, hope that the degeneracies do not prove problematic,
and adjust one's strategy to compensate if they do. One could argue
that fewer number of parameters would yield more predictive power.
However, if the parameter space considered does not contain the true
model nor a {}``signature'' model, the joint analysis may never
converge to physically meaningful result. The additional {}``predictive
power'' is then an illusion - in reality there is less predictive
power. Furthermore, the goal is not to rule out or rule in a particular
model, but rather to lay down a framework for looking for signatures
of classes of behavior (for example, the dark energy tracking the
other components). Given that we do not know what the dark energy
is, it would be unwise to risk incorrect assumptions implicit in the
choice of parameter space that artificially prevent interesting behavior
from being sampled. For example, if one used a parameter space limited
to minimally coupled scalar field models of dark energy, and the dark
energy had some characteristics consistent with this assumption (tracking)
and some inconsistent with this (present equation of state $<-1$)
one might falsely rule out the tracking - and in the end be entirely
mislead in understanding the underlying physics.

Another key aspect of ARDA is its flexibility: at a given moment a
variety of different analysis procedures can be performed (mixing
different observational datasets, parameter restrictions with various
theoretical priors) and it can also evolve over time to adapt to changes
in our understanding of cosmology and questions concerning the dark
energy (addition of new observational data, addition/removal of parameter
space dimensions, continually improving sampling, etc). The entire
parameter space need not be used in an analysis - one can instead
operate on theoretically-constrained subsurface. For example, a running
tensor index will not be strongly constrained alone. However, if one
assumes a theoretical framework that relates the running of the scalar
and tensor indexes, then one constrains the analysis to a sub-surface
of the full parameter space. Though no points will likely reside exactly
on this surface, Kernel Density Estimation gives a continuous, smooth,
analytic functional form for the likelihood function over the full
space. Every sub-space will therefore have an induced likelihood distribution
on it. This induced likelihood distribution will be used when theoretical
priors or constraints reduce the dimensionality of the full parameter
space. If the induced Kernel Density on a parameter subsurface does
not sufficiently sample the underlying distribution, the sampling
algorithm can be run on that subsurface alone. Also, as observations
improve and quantity and quality, more will be discovered about the
dark energy and the nature of the current questions will change. Some
parameters will be resolved to the extent that they can be fixed and
their range removed from the analysis, while some new parameters will
need to be added. A new dimension can easily be added as a new Importance
Sampling iteration begins by {}``puffing'' the distribution into
the new parameter dimension: one simply a-priori assigns a kernel
size centered on the previous assumed value of that parameter. Points
in the following iterations will drift into the new direction and
increasingly reflect the underlying distribution. In this way it is
expected that the ARDA system will never become redundant - instead
continuing to be an invaluable tool into the indefinite future.

The ARDA point dataset (parameter values plus cosmological {}``result''
of each point) will be made available to the scientific community.
Ultimately this will develop into an analysis software tool that will
allow the user to select an arbitrary set of experiments, and optionally
a subsurface of parameter space. As new experiments are conducted,
the experimental groups will be encouraged to provide their results
in a format suitable for inclusion into this analysis tool and shared
with the rest of the community. For an experiment to be added, one
only needs a function to be provided that maps a point in cosmological
parameter space (possibly using the associated CMB, power spectra,
etc) to an (unnormalized) likelihood for that experiment. The analysis
software can be distributed, or made available as a webpage. A working
proof-of-concept prototype, \underbar{The Cosmic Concordance Project}
can be viewed at \emph{\underbar{http://galadriel.astro.uiuc.edu/ccp/}}\emph{.}
The parameter space point set is several Markov chains that were created
to show how CMB data, with and without additional observational data
and a theoretical BBN constraint, can resolve the primordial abundance
of $He^{4}$~\cite{TR_GGHe4MCMC}. The parameter space and available
experiments are much more limited than ARDA, but this website demonstrates
the working concept for what will be developed. 

Additionally, a public outreach component is planned as a front-end
for the ARDA analysis webpage: a series of webpages will act as a
tutorial on cosmological parameter extraction for the interested general
public. A visitor will be able to read the tutorial to gain an understanding
of cosmology, the significance of the cosmological parameters, the
cosmological experiments and the statistical methods used to extract
allowed ranges of the parameters. Finally the visitor will be able
to use the actual parameter extraction website. This will educate
the general public about cosmology and cosmological parameter extraction,
as well as build an appreciation for experiments and encourage donation
of CPU time.

\section{The Experimental Component: Observational Data Modules}

A key feature of ARDA will be that it will allow likelihood functions
to be determined for an arbitrary set of experiments with little or
insignificant re-computational cost. In addition to performing parameter
extraction with every available dataset, the user will be able to
mix and match datasets. This is useful to determine the influence
of real and hypothetical experiments, as well as measure tension between
competing datasets. For example, apparent inconsistency between large
vs small scales, or early vs late time data could suggest that an
assumption of scale/time independence of a parameter is incorrect.
In such a case one could attempt to relax the tension by introducing
a new parameter to describe the scale/time variation. Such tests are
key to furthering our understanding of the dark sector, and it must
not be computationally prohibitive to perform a wide variety of such
tests with different parameters and parameter combinations. In the
ARDA system there will already be precomputed spectra and measured
quantities for each parameter space point. Thus when a dataset is
applied most of the computational work is already done. The author
envisions each experimental dataset available in ARDA will be a module
- at the core will be a likelihood functional which will take as input
cosmological parameters and associated precomputed spectra, and as
output produce an unnormalized likelihood for that experiment. The
ARDA user can make any choice of which experiment modules to use for
a given run. A working example of this modularity can been seen in
CosmoMC~\cite{CosmoMC} - it is distributed with some experimental
datasets included. The user can choose any of these as additional
weighting of the Markov chain point likelihood. Initially the dataset
modules will be created by the ARDA maintainers, but eventually it
is hoped that the research group running an experiment will provide
them.

\section{Conclusions}

ARDA will be an important framework for combining all cosmological
data to do the best possible job of parameter extraction - maximizing
what experimental data can teach us about our Universe, the dark energy
and the physics underlying it. Also, the hypothetical datasets of
a future experiment can be analyzed to determine how to maximize it's
impact in conjunction with existing data. Parameter confidence regions
can be determined for an arbitrary set of observational datasets without
significant recomputation cost. If the correct model of dark energy
is contained in the ARDA parameter space, improved data will tighten
the concordance region about it. Otherwise, as the data improves signatures
of the correct model will become apparent - such as tension between
classes of observations - which will direct us how to enlarge parameter
space to include the correct model. One possible sampling algorithm
to be used by ARDA, Importance Sampling/Kernel Density Estimation,
may prove to possess significant improvements over traditional Markov
chain algorithms in flexibility, parallelism and efficiency. ARDA
will be economical (new experiments can added with minimal computational
cost), general (experiments can be mixed and matched in an arbitrary
manner), extensible (new points can be added indefinitely), adaptable
(new parameters can be added), reducible (theoretical priors can be
used to study parameter constraints on lower dimensional subspaces)
and the parameter range is well motivated and physical (a wide range
of cosmological models are allowed). There is no reason for this analysis
tool to ever become obsolete - it can continue to be used indefinitely,
be adapted, and grow in predictive power as each new experiment is
incorporated. The scientific community can assist the development
of ARDA in two ways: by creating experiment modules, and by donating
CPU from otherwise idle computers. Interested parties should email
\underbar{$arda$$-devel$ $@$ isildur$.$astro$.$uiuc$.$edu}.

\begin{acknowledgments}
The author wishes to thank Benjamin Wandelt and David Larson for useful
discussions, and as collaborators in a more detailed treatment of
IS/KDE as a cosmological parameter extraction tool~\cite{KDEIS_for_CPE},
Antony Lewis, Anthony Challinor and Anthony Lasenby for useful discussions
and the CAMB code~\cite{CAMB}, Joe Mohr and Sarah Bridle for useful
discussions, and UIUC for support.
\end{acknowledgments}

\end{document}